\documentclass[a4paper,fleqn]{cas-dc}

\usepackage[numbers]{natbib}
\usepackage{algorithm}
\usepackage{algpseudocode}
\usepackage{float}
\usepackage{etoolbox}
\usepackage{graphicx}
\usepackage{float}
\usepackage{subfig}
\usepackage{amsmath}
\usepackage{amssymb}
\usepackage{caption}
\usepackage[numbers]{natbib}

\def\tsc#1{\csdef{#1}{\textsc{\lowercase{#1}}\xspace}}
\tsc{WGM}
\tsc{QE}
\tsc{EP}
\tsc{PMS}
\tsc{BEC}
\tsc{DE}

\begin{document}
\let\WriteBookmarks\relax
\def\floatpagepagefraction{1}
\def\textpagefraction{.001}
\shorttitle{Central Differential Flux with High-Order Dissipation}
\shortauthors{Bonan Xu et~al.}

\title [mode = title]{A Central Differential Flux with High-Order Dissipation for Robust Simulations of Transcritical Flows}                
\tnotemark[1,2]

\tnotetext[1]{This research is supported by the Start-up Fund for RAPs by the Hong Kong Polytechnic University.}


\author[1]{Bonan Xu}


\affiliation[1]{organization={Department of Aeronautical and Aviation Engineering, The Hong Kong Polytechnic University},
                addressline={11 Yuk Choi Road}, 
                city={Hong Kong SAR},
                country={China}}

\author[2]{Chang Sun}

\author[1]{Peixu Guo}
\cormark[1]
\ead{peixu.guo@outlook.com}


\affiliation[2]{organization={Department of International Cooperation, Beijing Microelectronics Technology Institute},
                addressline={No.2, North Siyingmen Road}, 
                postcode={100076}, 
                postcodesep={}, 
                city={Beijing},
                country={China}}

\cortext[cor1]{Corresponding author}


\begin{abstract}
The simulation of transcritical flows remains challenging due to strong thermodynamic nonlinearities that induce spurious pressure oscillations in conventional schemes.While primitive-variable formulations offer improved robustness under such conditions, they are always limited by energy conservation errors and the absence of systematic high-order treatments for numerical fluxes. In this paper, we introduce the Central Differential flux with High-Order Dissipation (CDHD), a novel numerical flux solver designed for primitive-variable discretization. This method combines a central flux for advection with a minimal, upwind-biased dissipation term to stabilize the simulation while maintaining formal accuracy. The dissipation term effectively suppresses oscillations and improves stability in transcritical flows. Compared to traditional primitive-variable approaches, CDHD reduces the energy conservation error in two order of magnitude. When incorporated into a hybrid framework with a conservative shock-capturing scheme, the method robustly handles both smooth transcritical phenomena and shock waves. Numerical tests validate the accuracy, stability, and energy-preserving capabilities of CDHD, demonstrating its potential as a reliable tool for complex real-gas flow simulations.
\end{abstract}



\begin{keywords}
transcritical flow \sep primitive-variable formulation \sep hybrid scheme \sep pressure oscillations
\end{keywords}

\maketitle

\section{Introduction}

\begin{figure}[h]
	\centering
	
	\subfloat[Density]{\includegraphics[scale=0.4]{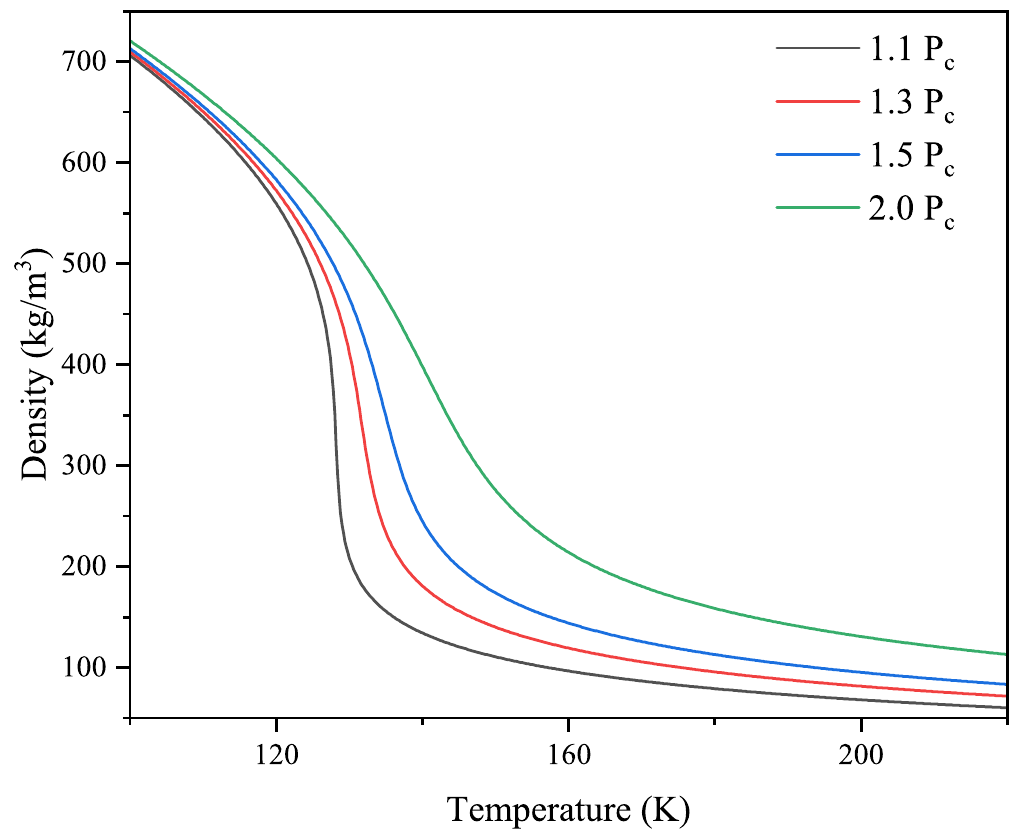}}\\
	\subfloat[Constant pressure specific heat capacity]{\includegraphics[scale=0.4]{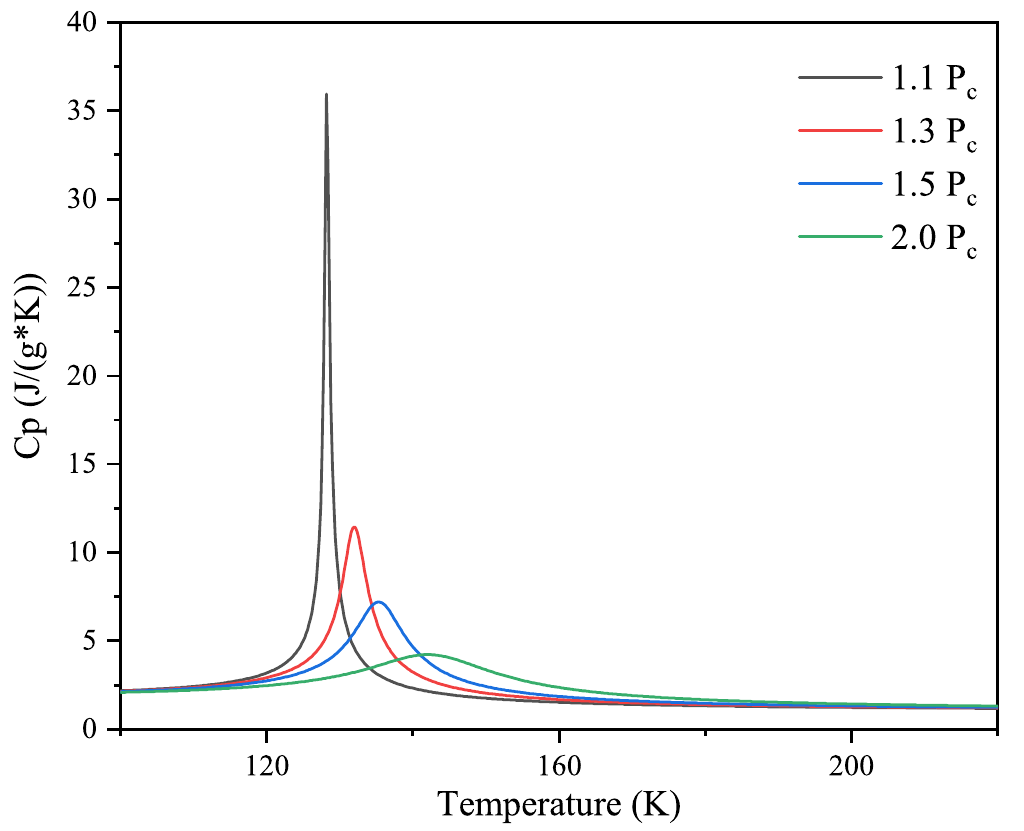}}
	\caption{Density and constant pressure specific heat capacity of transcritical nitrogen at different pressures. The critical pressure $P_c$ of nitrogen is $ 3.3958 \mathrm{MPa} $ and the critical temperature $T_c$ is $126.192 \mathrm{K}$.}
	\label{fig::fig1}
\end{figure}

The accurate simulation of fluids in the transcritical regime is essential for the design and analysis of advanced engineering systems\cite{annurev}, such as liquid rocket engines\cite{rocket1, rocket3, rocket2, rocket4}, supercritical power cycles\cite{power1, power2, power3}, and next-generation chip cooling systems\cite{chip1, chip2}. As shown in Figure~\ref{fig::fig1}, the regime located near the pseudo-boiling or Widom line\cite{Widom_line1,Widom_line2}, is characterized by drastic, liquid-like to gas-like variations in thermodynamic and transport properties. These strong thermodynamic nonlinearities introduce new fluid phenomena \cite{phys1, phys2, phy3, phy4} but present significant challenges for numerical simulations\cite{app1}. The application of widely used conservative schemes often generates severe unphysical pressure oscillations\cite{Ma_peter} that can render the simulation unstable. Even in ideal gas flows, similar oscillations can be observed in regions with rapid property change, such as material interfaces\cite{interfaces} or in multicomponent flows\cite{multicomponent1, multicomponent2, multicomponent3}. Thus, accurately capturing these transcritical phenomena requires not only a precise real-gas Equation of State (EoS), such as the Peng-Robinson\cite{PR1, PR2} model utilized in this paper, but also a numerical scheme that is both accurate and robust.

To address these numerical challenges, numerous attempts have been made to suppress these unphysical oscillations. For low-velocity transcritical flows, researchers have widely adopted the low-Mach-number approximation\cite{low_Mach_number1, low_Mach_number2, low_Mach_number3} of the governing equations. For instance, Pasquale\cite{low_Mach_number4} use a low-Mach approximation in the direct numerical simulation to fully resolve the interaction between turbulent and transcritical layers in a nitrogen jet. While effective and robust for low-velocity flows, this approximation can introduce non-negligible errors when the flow Mach number exceeds $0.1$. Thus, many new numerical methods have been developed based on non-conservative formulations. The double flux method\cite{double_flux} that is originally designed to reduce the pressure oscillation in multicomponent flows is used for the simulation of transcritical flows\cite{Ma_peter, double_flux1, double_flux2, double_flux3}. Instead of using the total energy equation, the pressure evolution equation\cite{pressure1, pressure2, pressure3, pressure4} is extensively employed for transcritical flow simulation. Kawai et al. successfully employ a high-order compact differencing scheme to solve the pressure evolution equation for a robust and accurate simulation of transcritical flows\cite{pressure4}. However, these non-conservative methods, including the double-flux and pressure evolution formulations, come at the cost of energy conservation error, which can lead to an incorrect prediction of shock wave locations and strengths. To reduce this conservation error, hybrid methods\cite{Hybrid1, Hybrid2, multicomponent3, Hybrid3} are also developed by researchers. Bradley Boyd and Dorrin Jarrahbashi \cite{Hybrid3} propose a hybrid scheme that switches between a double-flux in transcritical region and a traditional fully conservative method for other simulation domains. An emerging research topic in recent years is the development of fully conservative schemes to effectively suppress pressure oscillations during real-fluid simulations\cite{Conservative}. Eric Ching et at. develop a fully conservative discontinuous Galerkin method for supercritical, real-fluid flows \cite{L2proj} that uses an $L^{2}$  projection of primitive variables\cite{L2proj1, L2proj2, L2proj3} to reduce pressure oscillations. While this approach maintains stability, it can still result in small deviations from pressure equilibrium. More recently, Fujiwara et al. derive a pressure-equilibrium condition\cite{Conservative1} from the governing equations and constructed numerical fluxes to implicitly maintain this condition. Terashima et al. extend this method to real fluids\cite{Conservative2}, proposing an approximately pressure-equilibrium-preserving scheme with full energy conservation. However, these methods\cite{Conservative1, Conservative2, Conservative3} are not easily extended to general transcritical flows.

Despite these advances, the pressure evolution equation remains a promising approach for the robust simulation of transcritical flows. An open source code called CUBENS\cite{open_code} is recently developed based on the pressure evolution equation for the simulation of real fluid including transcritical flows. However, several key challenges persist. Firstly, discretizing the pressure evolution equation can lead to an incorrect prediction of shock wave locations and strength. Secondly, the energy conservation error should be minimized for industrial applications where heat transfer is critical. Finally, a robust, high-order, and easily extendable numerical flux solver for the primitive variable form of the equation has yet to be developed.

This paper presents a hybrid numerical method designed to address these challenges. We utilize a hybrid framework that switches between a non-conservative, primitive-variable update in smooth regions and a fully conservative update in shock regions. The Primitive Variable Riemann Solver (PVRS)\cite{toro} is utilized as a sensor to dynamically switch to a fully conservative scheme (fifth-order WENO\cite{WENO} with Roe flux\cite{Roe_real}) to ensure accurate and robust shock capturing. For the primitive-variable update, we propose a new interface flux solver called Central Differential flux with High-order Dissipation (CDHD) for the pressure evolution equation. This method uses a high-order central flux to provide accuracy for the convection term and augments it with a small, high-order, upwind-biased dissipation term to provide the necessary stabilization for transcritical gradients without degrading the formal order of accuracy. A rigorous error analysis shows that the overall order of accuracy is determined by the central flux, provided that the dissipation term is computed using high-order reconstructed values. Due to its low numerical diffusion, the proposed scheme demonstrates superior energy conservation in smooth regions compared to conventional primitive-variable methods. The hybrid framework ensures robustness for a wide range of flows containing both smooth transcritical phenomena and shocks.

The remainder of this paper is organized as follows: In Section~\ref{sec:math}, the mathematical model, including the governing equations and thermodynamic closure, is described. In Section~\ref{sec:numerical_method}, we detail the components of the hybrid numerical method. Section~\ref{sec:error} provides a formal error analysis justifying the accuracy of the proposed scheme. In Section~\ref{sec:result}, we present validation results from several one-dimensional (1D) and two-dimensional (2D) test cases to demonstrate the accuracy, stability, and robustness of the proposed scheme. Finally, Section~\ref{sec:conclusion} summarizes of the findings.

\section{Mathematical Model} \label{sec:math}
\subsection{Governing Equations}

The governing equations for inviscid transcritical fluid flows are the Euler equations. For clarity, the numerical method is derived and presented for the 1D case. The extension of the proposed method to multiple dimensions is straightforward, as will be demonstrated later with a two-dimensional test case.

The one-dimensional Euler equations can be expressed in a standard conservative form:

\begin{equation}
	\frac{\partial \mathbf{W}}{\partial t} + \frac{\partial \mathbf{F}(\mathbf{W})}{\partial x} = 0, 
\end{equation}
where $\mathbf{W} = [\rho, \rho u, E]^T$ represents the vector of conservative variables ($\rho$ for density, $u$ for velocity, $E$ for total energy), and $\mathbf{F}(\mathbf{W}) = [\rho u, \rho u^2 + p, u(E + p)]^T$ is the flux. The pressure $p$ is obtained from a real-gas EoS. Alternatively, in quasi-linear primitive form using $\mathbf{V} = [\rho, u, p]^T$, the equations become
\begin{equation}
	\frac{\partial \mathbf{V}}{\partial t} + B(\mathbf{V}) \frac{\partial \mathbf{V}}{\partial x} = 0, 
\end{equation}
where
\begin{equation}
	B(\mathbf{V}) = \begin{bmatrix}  
		u & \rho & 0 \\  
		0 & u & 1/\rho \\  
		0 & \rho a^2 & u  
	\end{bmatrix},
\end{equation}
and $a$ is the speed of sound, computed via the corresponding EoS of the working fluid.

\subsection{Thermodynamic Closure and Equation of State}

An accurate description of the thermodynamic nonlinearity for fluids in the transcritical regime is of primary significance. The extreme variations in fluid properties, particularly near the pseudo-boiling or Widom line, require that a thermodynamic model can accurately capture these strong nonlinearities. While the numerical method in this work is developed to be general and compatible with any sufficiently complex EoS, all computations herein will utilize the Peng-Robinson (PR) EoS\cite{PR1,PR2}. The PR-EoS is a cubic EoS that has been shown to reliably predict the thermodynamic properties\cite{EoS} for hydrocarbons and other fluids in transcritical and supercritical regions. Thus, the PR-EoS should be a suitable choice for the simulation of transcritical fluids.

To be specific, the PR EoS is expressed as
\begin{equation}
	P=\frac{\rho R_{u} T}{M_{w}-b \rho}+\frac{a \alpha(T) \rho^{2}}{M_{w}^{2}+2 M_{w} b \rho-b^{2} \rho^{2}},
\end{equation}
where $R_{u}$ is the universal gas constant, $M_{w}$ is the molar mass, and $T$ is the absolute temperature. The term $\alpha(T)$ is a temperature-dependent correction factor. The model parameters $a$ and $b$ are determined by the critical pressure $P_{c}$ and critical temperature $T_{c}$ of fluids\footnote{This $a$ is the parameter computed by the critical temperature and critical pressure of corresponding fluids for the cubic EoS, rather than speed of sound.}:

\begin{equation}
		a=0.45724\left(\frac{R_{u}^{2} T_{c}^{2}}{P_{c}}\right), \quad
		 b=0.07780\left(\frac{R_{u} T_{c}}{P_{c}}\right).
\end{equation}

The system of Euler equations is closed by defining the thermodynamic properties, such as the specific internal energy $e$ and the speed of sound $c$. These are derived from the chosen EoS using fundamental thermodynamic relations. For detailed derivations for real fluids, refer to the work of Kim et al\cite{EoS}.

The final expression for internal energy $e$ and the speed of sound squared $a^{2}$ are given by\cite{EoS}
\begin{equation}
	\begin{aligned}
		e(T, \rho) & =e_{0}(T)+\int_{\rho_{0}}^{\rho}\left[\frac{P}{\rho^{2}}-\frac{T}{\rho^{2}}\left(\frac{\partial P}{\partial T}\right)_{\rho}\right]_{T} d \rho, \\
		a^{2} & =\left(\frac{\partial P}{\partial \rho}\right)_{s}=\frac{c_{p}}{c_{v}}\left(\frac{\partial P}{\partial \rho}\right)_{T}.
	\end{aligned}
\end{equation}
Here, $e_{0}(T)$ refers to the internal energy of the reference ideal gas state, and the subscript $0$ denotes a reference ideal gas state, whose properties can be determined using standard equations, such as the NASA polynomials\cite{Ma_peter}. The terms $c_{p}$ and $c_{v}$ represent the specific heat capacities at constant pressure and constant volume, while the subscripts $s$ and $T$ denote isentropic and isothermal processes respectively.

\section{Numerical Method} \label{sec:numerical_method}

This work targets the robust simulation of real-fluid or transcritical compressible flows in which thermodynamic properties vary sharply across a small temperature range near the pseudo-boiling or Wisdom line. During the design of the algorithm for transcritical flows, there are two requirements: (i) strict conservation across shocks and contact discontinuities, and (ii) low spurious pressure-equilibrium (PE) errors when large thermodynamic gradients exist without true discontinuities. To meet both, we employ a hybrid scheme. Cells near the shock waves are updated in a conservative finite volume form using WENO-5 reconstruction\cite{WENO} and a Roe-type flux solver adapted to the real-gas EoS\cite{Roe_real}. Cells in the smooth region, transcritical region are updated in a primitive-variable form with a centered differential flux augmented by a small, upwind-biased dissipation. The primitive variables are reconstructed by WENO-5 to avoid characteristic projections through potentially ill-conditioned real-gas eigenstructures near the critical point. A conservative numerical scheme, such as the finite-volume method, is essential to ensure that the Rankine-Hugoniot jump conditions are satisfied\cite{hesthaven2017numerical}. This guarantees the correct prediction of the shock speed and strength, which is critical for physical accuracy.

\textbf{Notation:} Cell averages are denoted with the subscript $i$. Interface locations are at $x_{i \pm 1/2}$. Superscripts $+$ and $-$ always mean right-biased and left-biased one-sided limits at an interface (e.g., $V^{-}_{1+1/2}$ is the value reconstructed in the left side of cell $i+1/2$). Bold capital $\mathbf{W}$ is the vector of conservative variables, $\mathbf{V}$ is vector of primitive variables, and $B(\mathbf{V}) = \partial \mathbf{F}(\mathbf{V}) / \partial \mathbf{V}$ is the Jacobian of the inviscid flux written in primitive variables.

\subsection{Finite Volume Method}

The governing Euler equations in the integral conservative form for a one-dimensional control volume $[x_{i-1/2}, x_{i+1/2}]$ are given by\cite{hesthaven2017numerical}
\begin{equation}
	\begin{aligned}
		\frac{d}{d t} &  \int_{x_{i-1 / 2}}^{x_{i+1 / 2}} \mathbf{W}(x, t) d x \\ &+\mathbf{F}\left(\mathbf{W}\left(x_{i+1 / 2}, t\right)\right)-\mathbf{F}\left(\mathbf{W}\left(x_{i-1 / 2}, t\right)\right)=0,
	\end{aligned}
\end{equation}
where $\mathbf{W}$ is the vector of conserved variables and $\mathbf{F}$ is the  the corresponding flux vector. On a uniform grid of width $\Delta x$ and defining the cell-averaged value as $\mathbf{W}_{i}(t)=\frac{1}{\Delta x} \int_{x_{i-1 / 2}}^{x_{i+1 / 2}} \mathbf{W}(x, t) d x$, the equation becomes:
\begin{equation}
	\frac{d \mathbf{W}_{i}}{d t}=-\frac{1}{\Delta x}\left(\mathbf{F}_{i+1 / 2}-\mathbf{F}_{i-1 / 2}\right),
\end{equation}
where $\mathbf{F}_{i\pm 1 / 2}$ represents the numerical flux at the cell interfaces.

Using a forward Euler time integration (or a higher-order method like Runge-Kutta), the semi-discrete form with an explicit update can be written as
\begin{equation}
	\mathbf{W}{i}^{n+1}=\mathbf{W}{i}^{n}-\frac{\Delta t}{\Delta x}\left[\mathbf{F}_{i+\frac{1}{2}}-\mathbf{F}_{i-\frac{1}{2}}\right].
\end{equation}
The time step $\Delta t$ is controlled by the Courant-Friedrichs-Lewy (CFL) condition for stability:
\begin{equation}
	\Delta t = C_{\mathrm{cfl}} \frac{\Delta x}{S_{\max }^{(n)}},
\end{equation}
where $S_{\max }^{(n)}=\max _{i}\left\{\left|u_{i}^{n}\right|+a_{i}^{n}\right\}$ is the maximum characteristic wave speed in the domain at time level $n$. The CFL number $C_{\mathrm{cfl}}$ depends on the specific spatial and temporal schemes used.

\subsection{WENO Reconstruction}
In this work, we employ the WENO-5\cite{WENO} scheme to reconstruct the primitive variables directly. At each interface $x_{i+1/2}$, we compute right-biased ($v^{+}_{i+1/2}$) and left-biased ($v^{-}_{i+1/2}$) one side limits. In this subsection, superscripts $+$ and $-$ always mean right-biased and left-biased reconstruction at an interface.

To reconstruct a scalar component $v$ for the left biased value $v^{-}_{i+1/2}$, a 5-point stencil $\left\{v_{i-2}, v_{i-1}, v_{i}, v_{i+1}, v_{i+2}\right\}$ is used in WENO-5. This can be split into three 3-point sub-stencils:
\begin{equation}
	\begin{array}{l}
		S_{0}=\left\{v_{i-2}, v_{i-1}, v_{i}\right\} , \\
		S_{1}=\left\{v_{i-1}, v_{i}, v_{i+1}\right\} , \\
		S_{2}=\left\{v_{i}, v_{i+1}, v_{i+2}\right\} .
	\end{array}
\end{equation}
Each stencil is associated with a smoothness indicator ($\beta_0, \beta_1,\beta_2$), which quantifies the variation of variables within each stencil. These indicators are computed as follows:
\begin{equation*}
	\begin{aligned}
		\beta_{0}&=\frac{13}{12}\left(v_{i-2}-2 v_{i-1}+v_{i}\right)^{2}+\frac{1}{4}\left(v_{i-2}-4 v_{i-1}+3 v_{i}\right)^{2}, \\
		\beta_{1}&=\frac{13}{12}\left(v_{i-1}-2 v_{i}+v_{i+1}\right)^{2}+\frac{1}{4}\left(v_{i-1}-v_{i+1}\right)^{2}, \\
		\beta_{2}&=\frac{13}{12}\left(v_{i}-2 v_{i+1}+v_{i+2}\right)^{2}+\frac{1}{4}\left(3 v_{i}-4 v_{i+1}+v_{i+2}\right)^{2}.
	\end{aligned}
\end{equation*}

The raw weights $\alpha_k$ are determined by the smoothness indicators:
\begin{equation*}
	\alpha_{0}=\frac{d^{-}_{0}}{\left(\epsilon+\beta_{0}\right)^{2}}, \quad \alpha_{1}=\frac{d^{-}_{1}}{\left(\epsilon+\beta_{1}\right)^{2}}, \quad \alpha_{2}=\frac{d^{-}_{2}}{\left(\epsilon+\beta_{2}\right)^{2}},
\end{equation*}
where $\epsilon = 1 \times 10^{-6}$ is a small value added to avoid division by zero. The constants $d^{-}_{k}$ are the ideal weights that yield fifth-order accuracy in smooth regions. For the left-biased reconstruction $v^{-}_{i+1/2}$, these are $\left\{d_{0}^{-}, d_{1}^{-}, d_{2}^{-}\right\}=\{0.1,0.6,0.3\}$. The final normalized weights $\omega_k$ are then computed:

\begin{equation}
	\omega_{k}=\frac{\alpha_{k}}{\sum_{j=0}^{2} \alpha_{j}}.
\end{equation}
The reconstructed value is a weighted sum of polynomial approximations from each sub-stencil. 

A similar procedure is followed for the right-biased value at interface $v^{+}_{i+1/2}$, yet using a stencil shifted to the right $\left\{v_{i-1}, v_{i}, v_{i+1}, v_{i+2}, v_{i+3}\right\}$ and different ideal weights $\left\{d_{0}^{+}, d_{1}^{+}, d_{2}^{+}\right\}=\{0.3,0.6,0.1\}$.

\subsection{Shock Detection}

In this study, shock waves are detected using the PVRS\cite{toro}. In PVRS, the Riemann problem is approximately solved by:
\begin{align}
	P^{*} &= \frac{1}{2}\left(P_{L}+P_{R}\right)+\frac{1}{2}\left(u_{L}-u_{R}\right) \bar{\rho} \bar{a}, \\
	u^{*} & = \frac{1}{2}\left(u_{L}+u_{R}\right)+\frac{1}{2}\left(P_{L}-P_{R}\right) / (\bar{\rho} \bar{a}),\\
	\rho_{i+\frac{1}{2}}^{L}&=\rho_{i}+\left(u_{i}-u_{i+\frac{1}{2}}\right)(\bar{\rho} / \bar{a}), \\
	\rho_{i+\frac{1}{2}}^{R}&=\rho_{i+1}+\left(u_{i+\frac{1}{2}}-u_{i+1}\right)(\bar{\rho} / \bar{a}),
\end{align}
where $\bar{\rho}=\frac{1}{2}\left(\rho_{L}+\rho_{R}\right)$ and $\bar{a}=\frac{1}{2}\left(a_{L}+a_{R}\right)$ are the arithmetically averaged density and speed of sound at the interface, respectively.The propagation speeds of left and right waves are:
\begin{equation}
	\begin{aligned}
		s_{i+\frac{1}{2}}^{L}&=\frac{\left(\rho_{i} u_{i}-\rho_{i+\frac{1}{2}}^{L} u_{i+\frac{1}{2}}\right)}{\rho_{i}-\rho_{i+\frac{1}{2}}^{L}}, \\
		s_{i+\frac{1}{2}}^{R}&z=\frac{\left(\rho_{i+1} u_{i+1}-\rho_{i+\frac{1}{2}}^{R} u_{i+\frac{1}{2}}\right)}{\rho_{i+1}-\rho_{i+\frac{1}{2}}^{R}}.
	\end{aligned}
\end{equation}

Similar to previous research, the conservative methods are applied at cell $i$ when:
\begin{equation}
	\begin{aligned}
		& \frac{P_{i+\frac{1}{2}}}{P_{i}}>1+\varepsilon \quad \text { and } \quad s_{i+\frac{1}{2}}^{L}<0, \\
		\text { or } & \frac{P_{i-\frac{1}{2}}}{P_{i}}>1+\varepsilon \quad \text { and } \quad s_{i-\frac{1}{2}}^{R}<0,
	\end{aligned}
\end{equation}
where the parameter $\varepsilon$ can be selected within the range $(0, 0.1)$. In this paper, $\varepsilon = 0.05$ is used.

\subsection{Conservative Update}
For the cells near shock waves, a standard finite-volume WENO-5 scheme with Roe-type real-gas flux is used to update the conservative variable. Primitive variables $\mathbf{V}$ are reconstructed to obtain left ($\mathbf{V}_L = \mathbf{V}^+$) and right ($\mathbf{V}_R = \mathbf{V}^-$) states at interfaces, which are then converted to conservative variables $\mathbf{W}_L$ and $\mathbf{W}_R$. The Roe approximate Riemann solver designed for the real gas is used for the Roe flux. For more details, readers can refer to previous work\cite{Roe_real}. It is proved in previous research\cite{Hybrid1, Hybrid2, Hybrid3} that, compared with primitive form, the conservative form can maintain an accurate prediction of the speed and strength of shock waves.

\subsection{Primitive Update}

For cells in the smooth region, we discretize the non-conservative, quasi-linear form of the Euler equations:
\begin{equation}
	\frac{\partial \mathbf{V}}{\partial t}+\mathbf{B}(\mathbf{V}) \frac{\partial \mathbf{V}}{\partial x}=0,
\end{equation}
where $\mathbf{V}=[\rho, u, P]^{T}$ is the vector of primitive variables, and $\mathbf{B}(\mathbf{V})$ is the Jacobian matrix of primitive equations. The spatial term $\mathbf{L}_{v}=-\mathbf{B}(\mathbf{V}) \partial_{x} \mathbf{V}$ is handled by our proposed CDHD approach. This method split the flux into a central differential term and an upwind dissipative term:
\begin{equation}
	\mathbf{L}_{v}=\underbrace{\mathbf{L}_{v}^{\text {central }}}_{\text {accuracy }}+\underbrace{\mathbf{L}_{v}^{\text {diss }}}_{\text {stability }} .
\end{equation}
The central term is approximated using the WENO-5 reconstructed interface state:
\begin{equation}
	\begin{aligned}
		\left.\partial_{x} \mathbf{V}\right|_{i} &\approx \frac{\mathbf{V}_{i+1 / 2}^{-}-\mathbf{V}_{i-1 / 2}^{+}}{\Delta x},\\
		\mathbf{L}_{v}^{\text {central }}(i)&=-B\left(\mathbf{V}_{i}\right) \frac{\mathbf{V}_{i+1 / 2}^{-}-\mathbf{V}_{i-1 / 2}^{+}}{\Delta x}.
	\end{aligned}
\end{equation}

Given that this is a two-point centered difference on staggered locations, its truncation error is $O(\Delta x ^2)$ even though the individual interface states are fifth-order accurate. If desired, a higher order central derivative can be utilized without changing the rest of the formulation.

A small upwind-biased dissipation using path integrals in Dumbser–Osher–Toro Riemann solver (DOTRS)\cite{DOTRS1, DOTRS2} is added for stabilization:
\begin{equation}
	\begin{aligned}
		\mathbf{H}_{i+1 / 2}^{ \pm} &\approx\left(\int_{0}^{1} B^{\pm}(\Psi(s)) d s\right)\left(\mathbf{V}_{R}-\mathbf{V}_{L}\right), \\  \Psi(s)&=\mathbf{V}_{L}+s\left(\mathbf{V}_{R}-\mathbf{V}_{L}\right).
	\end{aligned}
\end{equation}
with $B^{\pm}=\frac{1}{2}(B\pm |B|)$ and a 3-point Gauss-Legendre rule for path integration. The dissipative contribution is then
\begin{equation}
	\mathbf{L}_{v}^{\mathrm{diss}}(i)=-\frac{1}{\Delta x}\left(\mathbf{H}_{i+1 / 2}^{-}+\mathbf{H}_{i-1 / 2}^{+}\right).
\end{equation}

In smooth regions, the jump $\mathbf{V}_{R}-\mathbf{V}_{L}=O\left(\Delta x^{5}\right)$ for WENO-5, so that $\mathbf{L}_{v}^{\mathrm{diss}}=O\left(\Delta x^{4}\right)$\footnote{For more details of the derivative process about the order of accuracy, please refer to the next section.}. Thus the dissipation is subdominant compared to the central term. In fact, the dissipation acts as a gentle upwind stabilization proportional to the spectral radius $|u|+a$, without degrading the formal order of the chosen central derivative.

\section{Error Analysis} \label{sec:error}
In this section, we provide a detailed mathematical justification for the behavior of the primitive update in smooth regions. We analyze the truncation error in the central term $\mathbf{L}_v^\text{central}$, which reveals a second-order accuracy, and the scaling of the state jump $d\mathbf{V} = \mathbf{V}_R - \mathbf{V}_L$, which leads to DOTRS fluctuations $\mathbf{H}^\pm = O(\Delta x^5)$ and a dissipative term $\mathbf{L}_v^\text{diss} = O(\Delta x^4)$. The dissipation magnitude is small enough not to influence the overall order of accuracy, namely the error from the central term dominates. We start by revisiting the key elements of WENO reconstruction and DOTRS, and then proceed with Taylor expansions and error analysis to reveal the behavior. By providing a detail error analysis, one can derive a higher order central flux easily.

\subsection{WENO-5 Reconstruction and State Jump Scaling}
The WENO-5 scheme, introduced by Jiang and Shu\cite{WENO}, reconstructs interface values from cell-averaged primitives $\mathbf{V}_i$ to achieve fifth-order accuracy in smooth regions. For a smooth scalar field $v(x)$, the cell average in cell $i$ (centered at $x_i$, with width $\Delta x$) is:
\begin{equation}
	\bar{v}_i = \frac{1}{\Delta x} \int_{x_{i-1/2}}^{x_{i+1/2}} v(\xi) \, d\xi = v(x_i) + O(\Delta x^2),
\end{equation}
where the Taylor expansion around $x_i$ shows that the approximation error is $O(\Delta x^2)$.

WENO5 computes left- ($v^{-}_{i+1/2}$) and right-biased ($v^{+}_{i+1/2}$) approximations at interface $x_{i+1/2}$ using a nonlinear weighted average of three candidate polynomials over stencils such that, in smooth regions, the combined reconstruction satisfies
\begin{equation}
	v^\pm_{i+1/2} = v(x_{i+1/2}) + O(\Delta x^5).
\end{equation}

At interface $i+1/2$ we set:
\begin{itemize}
	\item $V_L = V^+_i$: Right-biased reconstruction from the left cell $i$, approximating $V(x_{i+1/2}) + O(\Delta x^5)$.
	\item $V_R = V^-_{i+1}$: Left-biased from the right cell $i+1$, also approximating $V(x_{i+1/2}) + O(\Delta x^5)$.
\end{itemize}

In smooth regions, writing 
\begin{subequations}
	\begin{align}
		V_{L}&=V\left(x_{i+1 / 2}\right)+c_{L} \Delta x^{5}+O\left(\Delta x^{6}\right), \label{eq:Vla} \\
	    V_{R}&=V\left(x_{i+1 / 2}\right)+c_{R} \Delta x^{5}+O\left(\Delta x^{6}\right), \label{eq:Vlb}
	\end{align}
\end{subequations}
with (generally different) constants $c_{L}$ and $c_{R}$, the subtraction of equation~(\ref{eq:Vla}) and equation~(\ref{eq:Vlb}) yields $dV = (c_{R} - c_{L}) \Delta x^5 + O\left(\Delta x^{6}\right) = O\left(\Delta x^{5}\right)$. Lower-order terms are canceled because the scheme is provably fifth-order accurate at the interface. Thus, the jump is
\begin{equation}
	d V:=V_{R}-V_{L}=O\left(\Delta x^{5}\right).
\end{equation}

Near discontinuities the WENO nonlinearity lowers the order and $dV$ can be $O(1)$. Our hybrid method switches computations in such cells to conservative update, and thus the primitive update is used only where the $O\left(\Delta x^{5}\right)$ scaling holds.

\subsection{Central Flux Truncation Error}
We analyze the truncation error of the central part of the primitive update
\begin{equation} \label{eq:lvcentral}
	\mathbf{L}_{v}^{\text {central }}(i)=-B\left(\mathbf{V}_{i}\right) \frac{\mathbf{V}_{i+1 / 2}^{-}-\mathbf{V}_{i-1 / 2}^{+}}{\Delta x}
\end{equation}
in smooth regions. Two independent $O(\Delta x^2)$ mechanisms govern the accuracy: (1) the derivative approximation from two midpoint values, and (2) the use of $B\left(\mathbf{V}_{i}\right)$ (a cell averaged value) in place of $B\left(\mathbf{V}(x_{i})\right)$ (an accurate point value).

\subsubsection{Central difference from midpoints is second order}

Let $V$ be a smooth scalar component. By using exact interface values, one can arrive:

\begin{equation}
	\begin{aligned}
		&\frac{V\left(x_{i+1 / 2}\right)-V\left(x_{i-1 / 2}\right)}{\Delta x} \\
		&=V^{\prime}\left(x_{i}\right)+\frac{\Delta x^{2}}{24} V^{(3)}\left(x_{i}\right)+O\left(\Delta x^{4}\right).
	\end{aligned}
\end{equation}
Thus, the central-difference operator built from the two adjacent midpoints is inherently second order, regardless of how accurately those midpoint values are obtained.

When employing WENO-5 reconstruction in the smooth region, one observes
\begin{equation}
	V_{i \pm 1 / 2}^{ \pm}=V\left(x_{i \pm 1 / 2}\right)+O\left(\Delta x^{5}\right),
\end{equation}
so the reconstruction error contributes only
\begin{equation}
	\frac{\left[O\left(\Delta x^{5}\right)\right]-\left[O\left(\Delta x^{5}\right)\right]}{\Delta x}=O\left(\Delta x^{4}\right),
\end{equation}
which is subdominant to the $O(\Delta x^2)$ from the central difference itself.

\subsubsection{Coefficient mismatch of $B\left(\mathbf{V}_{i}\right)$ and $B\left(\mathbf{V}(x_{i})\right)$ is also $O(\Delta x^2)$}

$\mathbf{V}_{i}$ is a cell average $\mathbf{V}_{i}=\mathbf{V}\left(x_{i}\right)+O\left(\Delta x^{2}\right)$ by smoothing $B$:
\begin{equation}
	B\left(\mathbf{V}_{i}\right)=B\left(\mathbf{V}\left(x_{i}\right)\right)+O\left(\Delta x^{2}\right).
\end{equation}
By adding and subtracting the exact midpoint values in equation~(\ref{eq:lvcentral}), one can obtain
\begin{equation} \label{eq:central}
	\begin{aligned}
		&\mathbf{L}_{v}^{\text {central }}(i)=  -\underbrace{B\left(\mathbf{V}\left(x_{i}\right)\right)}_{\text {exact coeff. }} \underbrace{\frac{V\left(x_{i+1 / 2}\right)-V\left(x_{i-1 / 2}\right)}{\Delta x}}_{2 \text {-pt central diff }} \\
		& -\left(B\left(\mathbf{V}_{i}\right)-B\left(\mathbf{V}\left(x_{i}\right)\right)\right) \frac{\mathbf{V}_{i+1 / 2}^{-}-\mathbf{V}_{i-1 / 2}^{+}}{\Delta x} \\
		& -B\left(\mathbf{V}\left(x_{i}\right)\right)\left(\frac{\left.\mathbf{V}_{i+1 / 2}^{-}-\mathbf{V}_{i-1 / 2}^{+}-\frac{V\left(x_{i+1 / 2}\right)-V\left(x_{i-1 / 2}\right)}{\Delta x}\right)}{\Delta x}\right).
	\end{aligned}
\end{equation}
In the first line of the above equation~(\ref{eq:central}), the second-order central difference produces an $O(\Delta x^2)$ error. In the second line, $B\left(\mathbf{V}_{i}\right)-B\left(\mathbf{V}\left(x_{i}\right)\right)=O\left(\Delta x^{2}\right)$ times an $O(1)$ gradient also generates an $O(\Delta x^2)$ error. In the third line, reconstruction defects contribute only $O\left(\Delta x^{4}\right)$ error.Therefore, 
\begin{equation}
	\mathbf{L}_{v}^{\text {central }}(i)=-B\left(\mathbf{V}\left(x_{i}\right)\right) \partial_{x} \mathbf{V}\left(x_{i}\right)+O\left(\Delta x^{2}\right),
\end{equation}
so the overall truncation error of the central term is $O\left(\Delta x^{2}\right)$. Importantly, this second-order behavior is not due to WENO reconstruction (which is 5th order at the interfaces), but to (a) the two-midpoint central differencing and (b) the use of $B(\mathbf{V}_{i})$ evaluated at a cell average.

\subsection{Dissipative Term Scaling}

Let $d \mathbf{V}=\mathbf{V}_{R}-\mathbf{V}_{L}$ and the straight-line path $\Psi(s)=\mathbf{V}_{L}+s d \mathbf{V}, s \in[0,1]$. The DOTRS\cite{DOTRS1, DOTRS2} fluctuations are:
\begin{equation}
	\begin{aligned}
		\mathbf{H}^{ \pm}&=\int_{0}^{1} B^{ \pm}(\Psi(s)) d s d \mathbf{V},  \\
		\quad B^{ \pm}(\mathbf{V})&=\frac{1}{2}(B(\mathbf{V}) \pm|B(\mathbf{V})|),\\
		\quad|B|&=R|\Lambda| R^{-1},
	\end{aligned}
\end{equation}
with eigenvalues $\Lambda=\operatorname{diag}(u-c, u, u+c)$ and $\Psi(s)=\mathbf{V}_{L}+s d \mathbf{V}, s\in [0, 1]$.

Throughout this section we assume $B$ is $C^2$ (twice continuously differentiable) and that the eigen structure of $B(\mathbf V)$ is smooth along the path $\Psi(s)$ (no eigenvalue sign change), so that $|B|$ and $B^\pm=1/2(B\pm |B|)$ are differentiable.

Define $D$ as the Fréchet derivative (directional derivative of a matrix-valued function):
\begin{equation}
	D B^{ \pm}(\mathbf{V})[\delta \mathbf{V}]=\lim _{\tau \rightarrow 0} \frac{B^{ \pm}(\mathbf{V}+\tau \delta \mathbf{V})-B^{ \pm}(\mathbf{V})}{\tau}.
\end{equation}
The small-jump expansion can be derived as follows. Define $f(s)=B^{ \pm}(\Psi(s))$. By Taylor expansion in $s$ and the chain rule, one can arrive:
\begin{equation}
	\begin{aligned}
		f(s)=& B^{ \pm}\left(\mathbf{V}_{L}\right)+s D B^{ \pm}\left(\mathbf{V}_{L}\right)[d \mathbf{V}] \\
		&+\frac{1}{2} s^{2} D^{2} B^{ \pm}\left(\mathbf{V}_{L}\right)[d \mathbf{V}, d \mathbf{V}]+O\left(\|d \mathbf{V}\|^{3}\right).
	\end{aligned}
\end{equation}
By integrating the above equation, we obtain
\begin{equation} \label{eq:integrating}
	\begin{aligned}
		&\int_{0}^{1} f(s) d s = B^{ \pm}\left(\mathbf{V}_{L}\right) +\frac{1}{2} D B^{ \pm}\left(\mathbf{V}_{L}\right)[d \mathbf{V}] \\
		&+\frac{1}{6} D^{2} B^{ \pm}\left(\mathbf{V}_{L}\right)[d \mathbf{V}, d \mathbf{V}]+O\left(\|d \mathbf{V}\|^{3}\right).
	\end{aligned}
\end{equation}
Multiplying equation~(\ref{eq:integrating}) by $d\mathbf{V}$ gives
\begin{equation}
	\mathbf{H}^{ \pm}=B^{ \pm}\left(\mathbf{V}_{L}\right) d \mathbf{V}+\frac{1}{2} D B^{ \pm}\left(\mathbf{V}_{L}\right)[d \mathbf{V}] d \mathbf{V}+O\left(\|d \mathbf{V}\|^{3}\right).
\end{equation}
Hence, $\left\|\mathbf{H}^{ \pm}\right\| \leq C_{1}\|d \mathbf{V}\|+C_{2}\|d \mathbf{V}\|^{2}$ , i.e. linear in $d\mathbf{V}$ to the leading order.

By using WENO-5 reconstruction, the reconstructed interface states in smooth region satisfy:
\begin{equation}
	\begin{aligned}
		\mathbf{V}_{L} & =\mathbf{V}\left(x_{i+1 / 2}\right)+O\left(\Delta x^{5}\right), \\
		\mathbf{V}_{R} & =\mathbf{V}\left(x_{i+1 / 2}\right)+O\left(\Delta x^{5}\right).
	\end{aligned}
\end{equation}
Thus, $d \mathbf{V}=O\left(\Delta x^{5}\right)$. With the operator norm being controlled by $\| \mathbf{B}\left(\mathbf{V}_\text {avg }\right) \| \lesssim C(|u|+c)$, one can arrive
\begin{equation}
	\begin{aligned}
		\mathbf{H}^{ \pm}&=O\left(\Delta x^{5}\right) \quad \text { and }  \\
		\mathbf{L}_{v}^{\text {diss }}(i)&=-\frac{1}{\Delta x}\left(\mathbf{H}_{i+1 / 2}^{-}+\mathbf{H}_{i-1 / 2}^{+}\right)=O\left(\Delta x^{4}\right) .
	\end{aligned}
\end{equation}
Thus the dissipation is high-order small and does not limit accuracy when combined with the central term. The total $\mathbf{L}_v = \mathbf{L}_v^\text{central} + \mathbf{L}_v^\text{diss}$ thus has an overall truncation error $O(\Delta x^2)$ in smooth regions, remaining spatial accuracy to the second order.

\subsection{The Role of DOTRS in First-Order vs. High-Order Schemes}
The function of the DOTRS dissipation changes dramatically with the order of the underlying scheme.

\subsubsection{Role in First-Order Schemes}

In a first-order Godunov-type scheme, states are assumed to be piecewise constant, so $\mathbf{V}_{L} = \mathbf{V}_{i}$ and $\mathbf{V}_{R} = \mathbf{V}_{i+1}$. Jumps across interfaces are large ($d \mathbf{V}=O(1)$). Here, the DOTRS integral acts as the primary Riemann solver, providing the full upwind flux necessary for stability and physical resolution. The resulting spatial discretization  $L_v(i) = -\frac{1}{\Delta x} (H^-_{i+1/2} + H^+_{i-1/2})$ is first-order accurate.

\subsubsection{Role in This High-Order Scheme}
In our high-order method, the primary advection is handled by the accurate central flux term. The role of DOTRS is reduced to that of a targeted stabilizer. Given that high-order reconstruction ensures the interface jump $d\mathbf{V}$ is very small ($O(\Delta x^5)$)  in smooth regions, the DOTRS fluctuation $\mathbf{H}^{\pm}$ also becomes minor. For a small $d\mathbf{V}$, the dissipation behaves similarly to a local Lax-Friedrichs scheme, whereas with a crucial difference that its magnitude is scaled by the high-order jump. The resulting dissipative term in the final update $\mathbf{L}_{v}^{\text {diss}}$ is $O(\Delta x^4)$. This dissipation is numerically small enough such that it vanishes faster than the error of the second-order central term as the grid is refined. Therefore, it successfully provides stabilization against oscillations without affecting the formal second-order accuracy of the overall scheme.

\section{Result and Discussion} \label{sec:result}
In this section, a series of $1\text{D}$ and $2\text{D}$ numerical tests are undertaken to illustrate the accuracy and stability of the proposed scheme. The transcritical nitrogen is adopted as the working fluid in all of the test cases with the thermodynamic properties of fluids obtained by PR EoS. The numerical results are compared with available standard solutions. For a low magnitude dissipation, the WENO-5 reconstruction is used in all cases. The following cases show that the numerical scheme proposed in this paper can achieve an accurate, oscillation free simulation in transcritical regions.

\subsection{$1\text{-D}$ Transcritical Advection Cases}

\subsubsection{Result of CDHD Scheme}
A one-dimensional transcritical advection case is considered in a periodic domain $\Omega \in [0, 1]$ m. The initial conditions consist of a sharp but continuous temperature gradient centered within the domain, while the velocity and pressure fields are spatially uniform. The setup mimics a droplet in a transcritical environment, which is defined as
\begin{equation}
	\begin{aligned}
		&T(x)=T_{l} \\
		&+\frac{T_{r}-T_{l}}{2}\left[\tanh \left(\frac{x-0.25}{\eta}\right)+\tanh \left(\frac{-(x-0.75)}{\eta}\right)\right], \\
		&u(x)=100 \quad(\mathrm{~m} / \mathrm{s}), \qquad p(x)=4.0 \times 10^{6} \quad(\mathrm{~Pa}),
	\end{aligned}
\end{equation}
where, $T_L = 200$ K and $T_R = 100$ K are the left and right reference temperatures, respectively; $\eta = 0.1$ is a smoothing parameter controlling the width of the temperature transition. The density field $\rho(x)$ is computed using the PR EoS.

\begin{figure}
	\centering
	\subfloat[Density]{\includegraphics[scale=0.3]{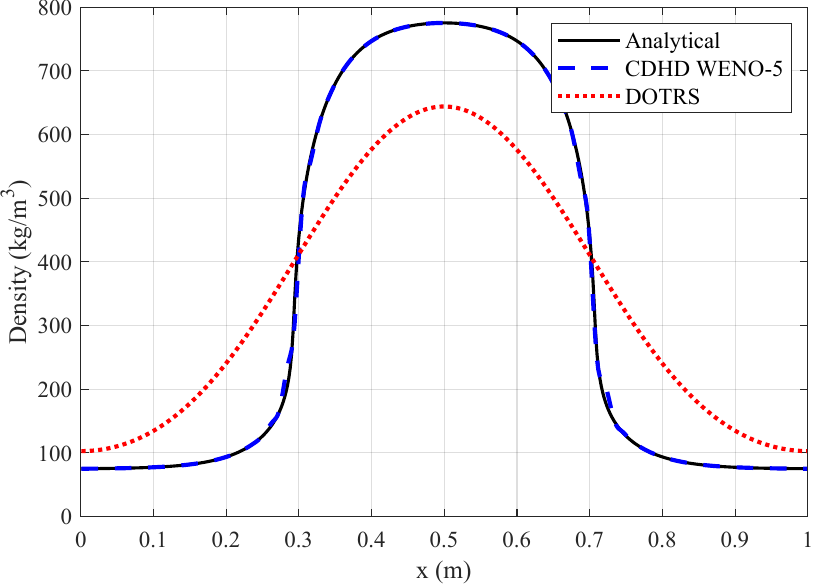}}
	\subfloat[Velocity]{\includegraphics[scale=0.3]{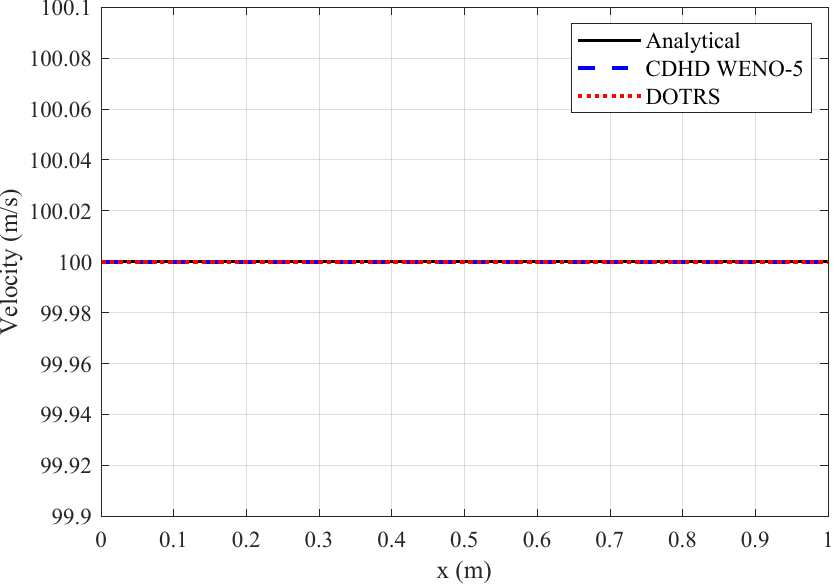}}
	\\
	\subfloat[Pressure]{\includegraphics[scale = 0.3]{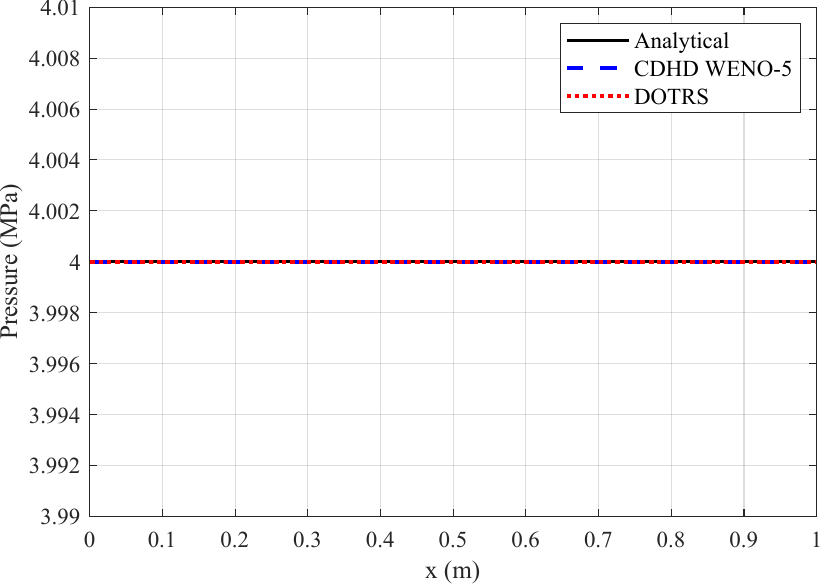}}
	\subfloat[Temperature]{\includegraphics[scale = 0.3]{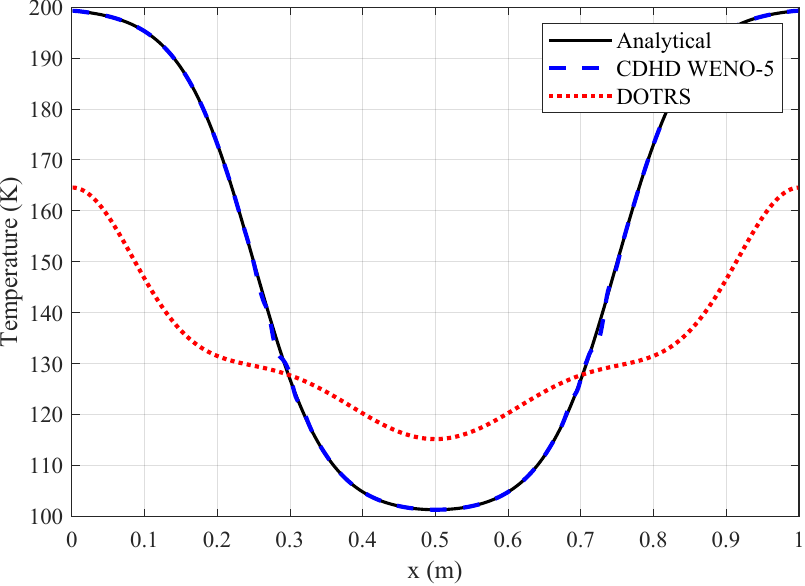}}
	\caption{Comparison of numerical results from the CDHD WENO-5 and the first-order DOTRS schemes with the analytical solution for the 1-D transcritical advection case at $t=0.1$ s (ten domain traversals) using $512$ grid points. Profiles of (a) density, (b) velocity, (c) pressure, and (d) temperature are shown. The CDHD WENO-5 scheme shows an excellent agreement with the analytical solution, while the DOTRS scheme exhibits significant numerical smearing in the density and temperature profile.}
	\label{fig::1dcp}
\end{figure}

The proposed CDHD method with a WENO-5 reconstruction is validated against a transcritical convection problem with a known analytical solution. The numerical results at the simulation time $t=0.1 \mathrm{\ s}$, corresponding to $10$ periods of convection, are presented in Figure~\ref{fig::1dcp}. All results are simulated with $512$ grid points.

For the velocity and pressure fields, both the CDHD WENO-5 and the first-order DOTRS schemes demonstrate an excellent agreement with the analytical solution. Both methods achieve a robust simulation, accurately maintaining the constant velocity of $100\mathrm{\ m/s}$ and pressure of $4.0 \mathrm{\ MPa}$ across the entire spatial domain. However, significant differences emerge in the temperature and density profiles. The density field exhibits more rapid variations due to the nonlinear thermodynamic effect near the critical point. The CDHD WENO-5 method accurately reproduces both the density and temperature profiles, preserving the sharpness of transitions without introducing spurious oscillations. In contrast, the first-order DOTRS scheme results in noticeable smearing of the interface, limiting its usage in resolving sharp gradients.

Overall, the CDHD method with WENO-5 reconstruction demonstrates superior accuracy and robustness in simulating transcritical flows, particularly in capturing the nonlinearities arising from thermodynamic properties. The method delivers robust performance across all evaluated fields including density, velocity, pressure, and temperature. Our work provides a new strategy for the further development of the high-order scheme for transcritical flows.

\begin{figure}
	{\includegraphics[scale=0.45]{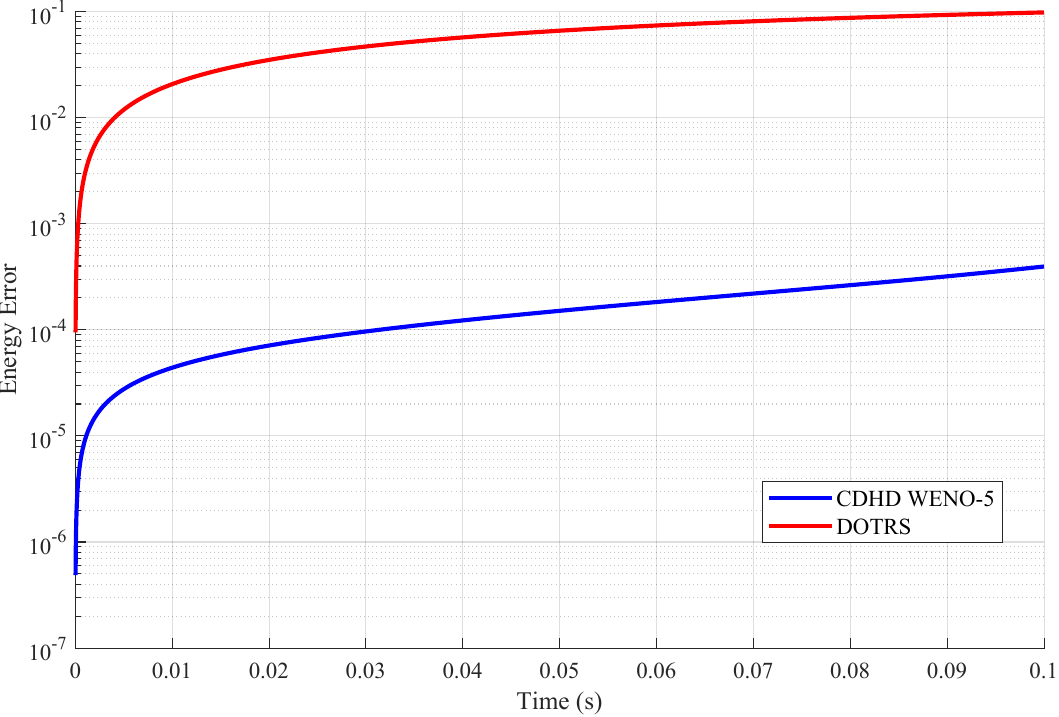}}
	\caption{Temporal evolution of the relative energy conservation error for the CDHD WENO-5 and the first-order DOTRS schemes, plotted on a semi-logarithmic scale. The error for the proposed CDHD WENO-5 scheme remains below $0.04\%$ at the final time of $t=0.1 \mathrm{\ s}$, while the error for the DOTRS scheme grows to approximately $10\%$ at the same instant.}
	\label{fig:interface}
\end{figure}

\subsubsection{Energy Conservation Error}
The primitive method has long been criticized for violating the law of energy conservation. However, the results from the proposed CDHD WENO-5 scheme demonstrate substantial improvements in energy conservation. We further examine the temporal evolution of the relative energy conservation error, defined as $\epsilon_E = \left| \frac{E(t) - E(0)}{E(0)} \right|$, where $E(t)$ is the integrated total energy at time $t$. Figure~\ref{fig:interface} presents this error metric over the simulation duration from $t = 0$ to $t=0.1\mathrm{s}$ (equivalent to $10$ convection periods) on a semi-logarithmic scale.

The results clearly demonstrate the superior performance of the proposed method in conserving energy. The CDHD WENO-5 scheme maintains a low energy error throughout the simulation, reaching a final error value of approximately $3.95 \times 10^{-4}$ at $t = 0.1 \mathrm{\ s}$. This corresponds to an error of less than $0.04 \  \%$, indicating a high degree of energy conservation suitable for long-duration simulations. In contrast, the first-order DOTRS scheme exhibits a significantly larger error that grows to about $10 \ \%$ by the final time point. This direct comparison reveals that the energy conservation error of the CDHD WENO-5 scheme is over two orders of magnitude lower than that of the DOTRS scheme at the end of the simulation.

\begin{table*}[htbp]
	\centering
	\caption{Convergence study results for L1-Error.}
	\label{tab:convergence_simple}
	\begin{tabular}{l c c c c c c c}
		\hline
		\textbf{$N_x$} & \textbf{$dx$} & \textbf{$L_1$-Error($\rho$)} & \textbf{Order} & \textbf{$L_1$-Error($\rho u$)} & \textbf{Order} & \textbf{$L_1$-Error($E$)} & \textbf{Order} \\
		\hline
		32   & $3.1250 \times 10^{-2}$ & $1.7647 \times 10^{1}$ & N/A    & $1.7647 \times 10^{3}$ & N/A    & $6.9015 \times 10^{6}$ & N/A    \\
		64   & $1.5625 \times 10^{-2}$ & $8.3227 \times 10^{0}$ & 1.0843 & $8.3227 \times 10^{2}$ & 1.0843 & $3.1120 \times 10^{6}$ & 1.1490 \\
		128  & $7.8125 \times 10^{-3}$ & $2.6924 \times 10^{0}$ & 1.6281 & $2.6924 \times 10^{2}$ & 1.6281 & $9.9529 \times 10^{5}$ & 1.6447 \\
		256  & $3.9062 \times 10^{-3}$ & $9.5399 \times 10^{-1}$ & 1.4969 & $9.5399 \times 10^{1}$ & 1.4969 & $3.4484 \times 10^{5}$ & 1.5292 \\
		512  & $1.9531 \times 10^{-3}$ & $2.6181 \times 10^{-1}$ & 1.8654 & $2.6181 \times 10^{1}$ & 1.8654 & $9.2446 \times 10^{4}$ & 1.8993 \\
		1024 & $9.7656 \times 10^{-4}$ & $5.8015 \times 10^{-2}$ & 2.1740 & $5.8015 \times 10^{0}$ & 2.1740 & $2.0015 \times 10^{4}$ & 2.2075 \\
		\hline
	\end{tabular}
\end{table*}

\begin{figure}
	\centering
	{\includegraphics[scale=0.5]{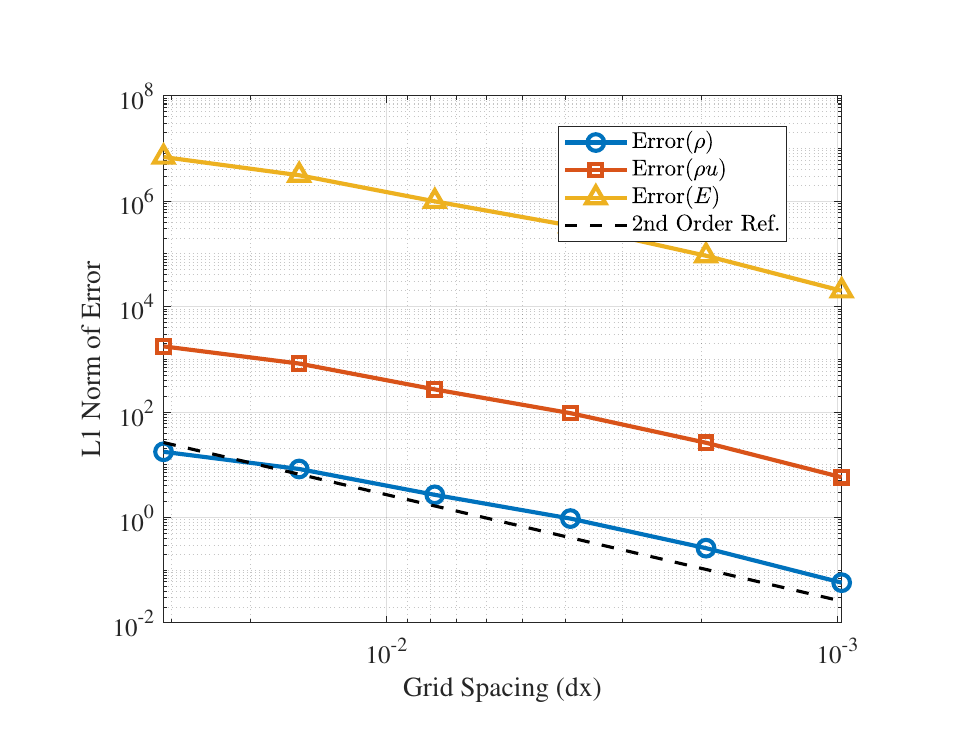}}
	\caption{The L1 norm of the error is plotted against grid spacing ($dx$) to demonstrate the order of accuracy of the numerical scheme. The convergence rates for density, momentum, and total energy are shown. A reference line with a slope of two is plotted to represent a second-order accurate scheme.}
	\label{fig:convergence}
\end{figure}

\subsubsection{Convergence Test}
Table~\ref{tab:convergence_simple} and Figure~\ref{fig:convergence} present the $L_1$ errors for $(\rho,\rho u,E)$ under uniform mesh refinement. While the two coarsest grids exhibit pre-asymptotic behavior, the method achieves near-second-order convergence for $N_x=256 \to 1024$. Specifically, least-squares fits over the three finest meshes yield convergence rates of $r=(2.17,2.17,2.21)$ for $(\rho,\rho u,E)$, which aligns well with our theoretical error analysis.

This second-order accuracy results from the careful balance of several algorithmic components. The WENO-5 reconstruction delivers fifth-order interface states, while the primitive-variable central difference scheme limits the smooth-region accuracy to $O(\Delta x^2)$. Crucially, the DOTRS term contributes only $O(\Delta x^4)$ dissipation, effectively stabilizing real-gas gradients without degrading the leading-order accuracy. For applications requiring higher-order accuracy in smooth regions, the overall scheme order can be increased by replacing the central gradient with higher-order central or split-flux differencing.

\begin{figure}
	\centering
	\subfloat[Density]{\includegraphics[scale=0.3]{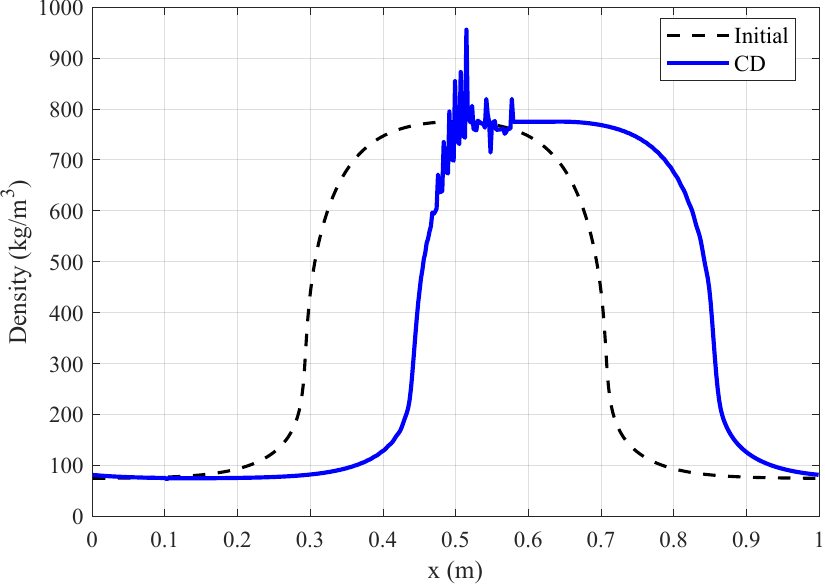}}
	\subfloat[Velocity]{\includegraphics[scale=0.3]{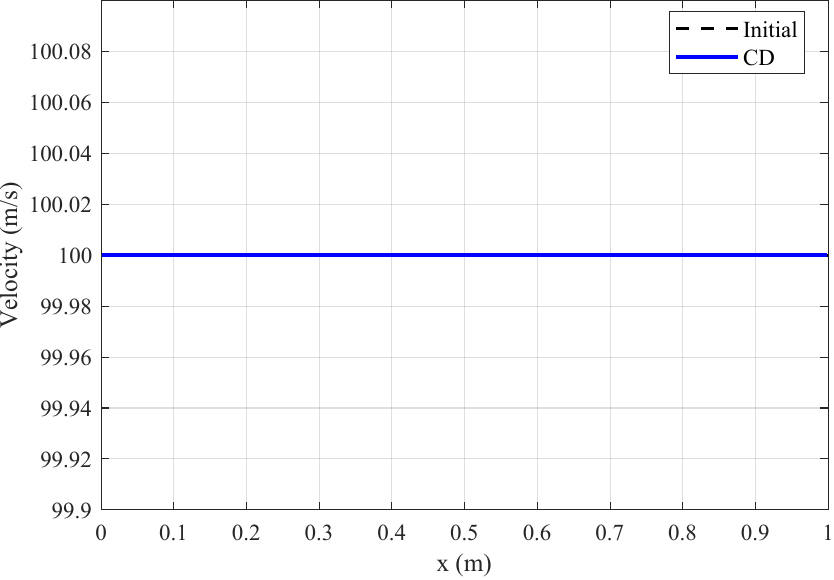}}
	\\
	\subfloat[Pressure]{\includegraphics[scale = 0.3]{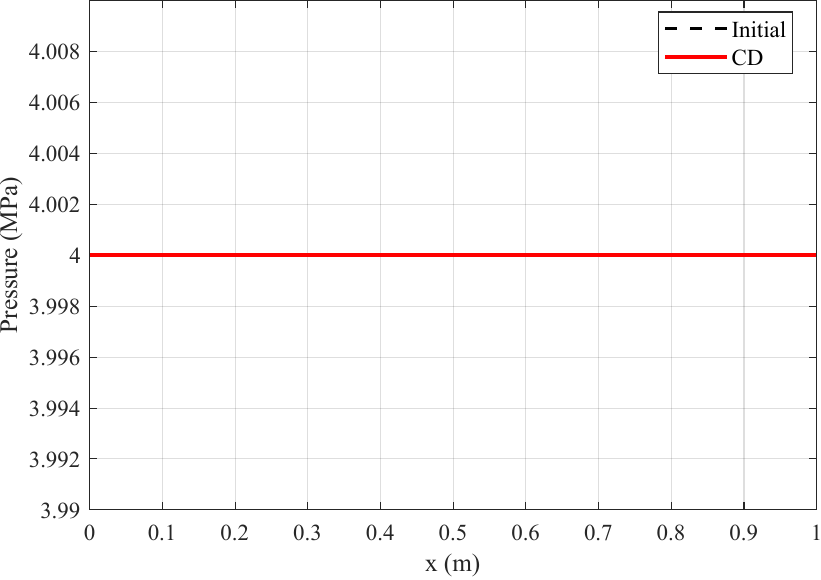}}
	\subfloat[Temperature]{\includegraphics[scale = 0.3]{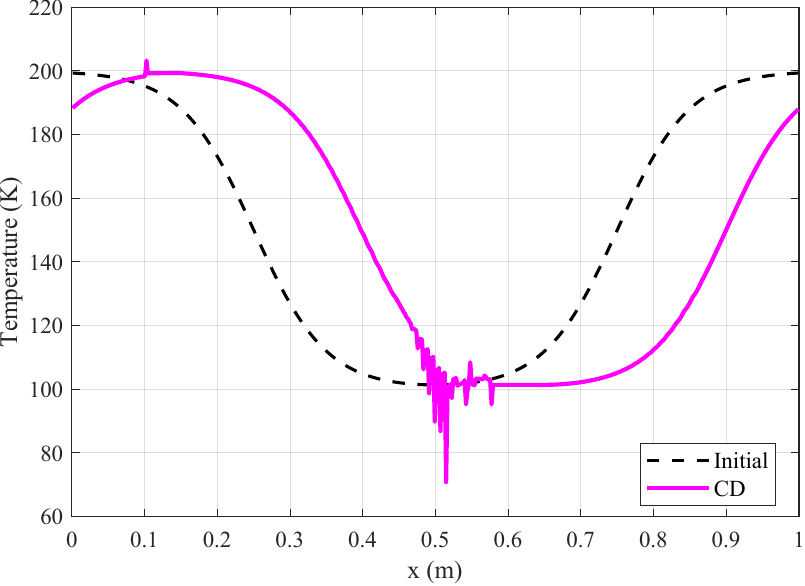}}
	\caption{Profiles of density, velocity, pressure and temperature for 1-D transcritical advection case when only using the central differential interface flux. Oscillation will be accumulated in time and blow up the simulation quickly.}
	\label{fig::1DCD}
\end{figure}

\subsubsection{Role of High-order Dissipation}

To isolate the effect of the dissipative term, we repeat the transcritical advection test case using only the central differential flux for $\partial_x\mathbf{V}$, omitting the high-order dissipative stabilization. The results in Figures~\ref{fig::1DCD} after $600$ time steps with $\text{CFL}=0.8$ clearly demonstrate the necessity of dissipation in the transcritical regime. The small magnitude dissipation proposed in this paper can greatly stabilize the simulation process.

As shown in Figure~\ref{fig::1DCD}, the central difference scheme without stabilization develops severe numerical instabilities. The density field exhibits spurious oscillations and overshoots near the steep gradient region, with spurious wiggles corrupting the solution throughout the domain. Similarly, the temperature field displays significant nonphysical oscillations that grow rapidly in time. In contrast, the velocity and pressure fields remain relatively stable, staying close to their initial uniform states with only minor perturbations.This selective instability pattern is characteristic of transcritical flow physics, where small perturbations in thermodynamic properties can trigger rapid growth of numerical modes.

These results demonstrate that a small amount of high-order, characteristic-based dissipation is essential for stabilizing the primitive variable update in transcritical flows. The dissipative term proposed in this work provides this stabilization without compromising the leading-order accuracy of the scheme, as evidenced by the convergence analysis presented earlier.

\begin{figure}
	\centering
	\subfloat[Density]{\includegraphics[scale=0.3]{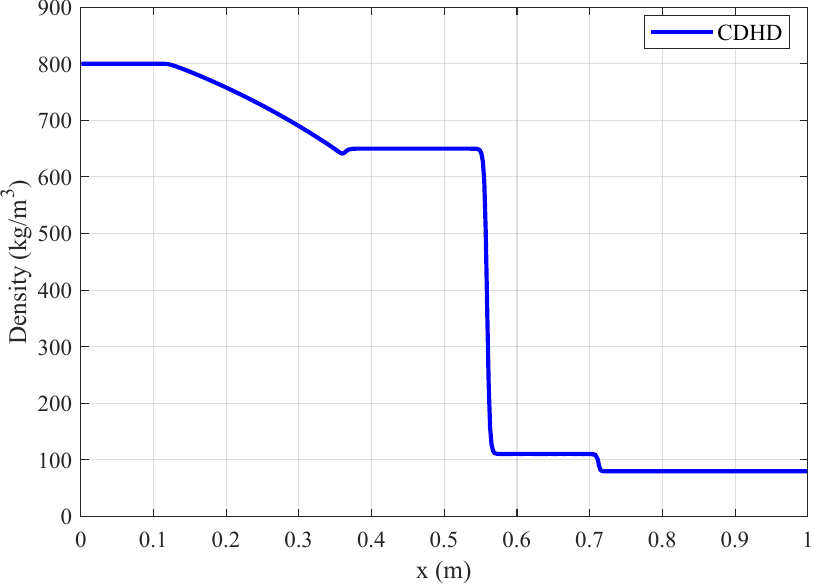}}
	\subfloat[Velocity]{\includegraphics[scale=0.3]{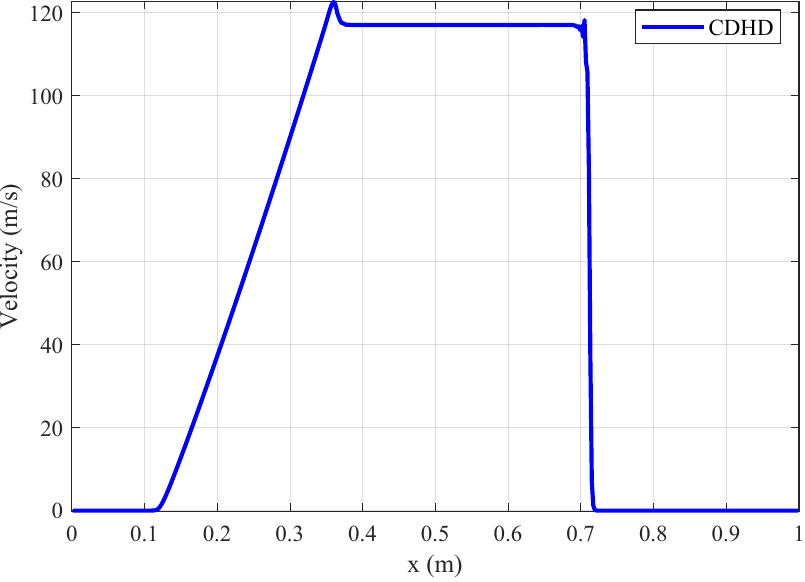}}
	\\
	\subfloat[Pressure]{\includegraphics[scale = 0.3]{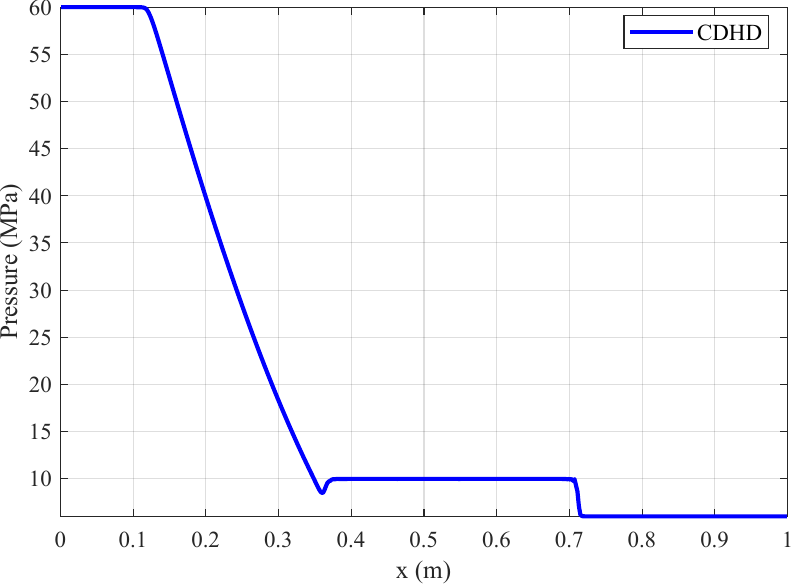}}
	\subfloat[Temperature]{\includegraphics[scale = 0.3]{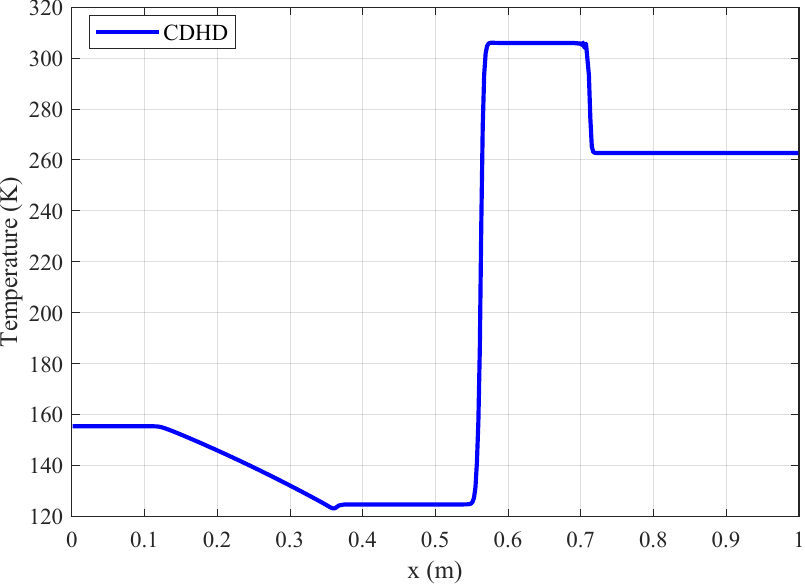}}
	\caption{Profiles of density, velocity, pressure and temperature for 1-D shock tube problems.}
	\label{fig::1Dshock}
\end{figure}

\subsection{$1\text{-D}$ Shock Tube Problems}

This section demonstrates the capability of the hybrid scheme in this work for accurately capturing the strength and propagation speed of shock waves through a $1\text{-D}$ shock tube test case. We consider a computational domain of $x \in [0, 1]$ m with transmissive boundary conditions applied at both ends. All simulations employ a CFL number $0.8$ and are run until $t = 5 \times 10^{-4}$ s. The initial conditions for the transcritical shock tube problem are specified as a Riemann problem with left and right states:
\begin{equation}
	\begin{array}{l}
		{\left[\begin{array}{c}
				\rho_{l} \\
				u_{l} \\
				p_{l}
			\end{array}\right]=\left[\begin{array}{c}
				800 \mathrm{~kg} / \mathrm{m}^{3} \\
				0 \mathrm{~m} / \mathrm{s} \\
				60 e 6 \mathrm{~Pa}
			\end{array}\right],} \\ \\
		{\left[\begin{array}{c}
				\rho_{r} \\
				u_{r} \\
				p_{r}
			\end{array}\right]=\left[\begin{array}{c}
				80 \mathrm{~kg} / \mathrm{m}^{3} \\
				0 \mathrm{~m} / \mathrm{s} \\
				6 e 6 \mathrm{~Pa}
			\end{array}\right]}.
	\end{array}
\end{equation}
This configuration creates a strong pressure and density ratio of the left and right sides in the simulation domain, establishing conditions that span the transcritical regime and challenge numerical methods with steep gradients and rapid thermodynamic variations.

Figure~\ref{fig::1Dshock} presents the computed density, velocity, pressure, and temperature profiles without using any limiter or filter. The solutions demonstrate the adaptive scheme has the ability to resolve the complex wave structure typical of transcritical shock tubes, including shock waves, contact discontinuities, and expansion regions, while maintaining sharp interfaces without spurious oscillations.

\begin{figure}
	\centering
	{\includegraphics[scale=0.6]{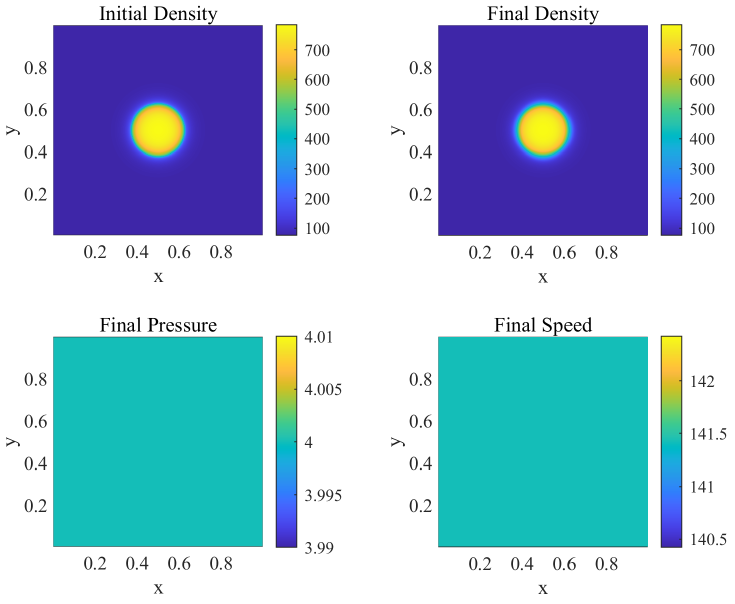}}
	\caption{Results for the 2D transcritical droplet advection case at the final time of $t=0.01 \mathrm{\ s}$ (one period). The figure displays the initial and final density fields, along with the final pressure and speed. The droplet maintains its circular shape with minimal numerical diffusion, while the pressure and speed fields remain uniform, demonstrating the accuracy of the proposed scheme.}
	\label{fig:2D}
\end{figure}

\subsection{$2\text{-D}$ Transcritical Advection Cases}

To demonstrate the capability of the proposed scheme in multi-dimensional transcritical flows, we consider a 2D droplet convection problem that extends the 1D analysis to more realistic geometries. This test case examines the transport of a subcritical droplet through a supercritical environment under uniform advection.

The computational domain is defined as $\Omega \in [0, 1] \times [0, 1]$ m with a uniform Cartesian grid of $151 \times 151$ cells. The initial state is a circular droplet centered at $(x_c, y_c) = (0.5, 0.5)$ m with radius $r_d = 0.15$ m. The temperature field is initialized using a smooth hyperbolic tangent profile to represent the droplet interface:
\begin{equation}
	\begin{aligned}
		&T(x,y) = T_{\text{ambient}} \\
		 &+ (T_{\text{droplet}} - T_{\text{ambient}}) \cdot \frac{1}{2}\left[1 - \tanh\left(\frac{r - r_d}{\eta}\right)\right],
	\end{aligned}
\end{equation}
where $r = \sqrt{(x - x_c)^2 + (y - y_c)^2}$ is the radial distance from the droplet center, $T_{\text{droplet}} = 100$ K represents the subcritical liquid phase, $T_{\text{ambient}} = 200$ K corresponds to the supercritical gas phase, and $\eta = 0.05$ controls the interface thickness. The velocity and pressure fields are initialized as spatially uniform:
\begin{equation}
	\mathbf{u}(x,y) = (100, 100)^T \text{ m/s}, \quad p(x,y) = 4.0 \times 10^6 \text{ Pa}.
\end{equation}
The density field $\rho(x,y)$ is computed from the temperature and pressure using the PR EoS. The simulation runs until $t = 0.01 \mathrm{s}$ with $\text{CFL}=0.5$, corresponding to one complete convection time. This set up allows the initial state to serve as a analytical solution.

Figure~\ref*{fig:2D} presents the evolution of the transcritical droplet convection, showing the initial and final density distributions along with the final pressure and velocity magnitude fields. The density field evolution reveals that the droplet maintains its circular shape and sharp interface throughout the convection process. The initial density contrast between the subcritical droplet core and the supercritical ambient medium is preserved without significant numerical diffusion. The absence of spurious oscillations near the steep density gradients confirms the efficacy of the WENO-5 reconstruction in handling transcritical interfaces. The pressure and velocity fields exhibit uniformity across the entire domain. The proposed scheme successfully maintains pressure equilibrium while resolving the dramatic thermodynamic variations at the droplet interface. These results confirm that the high-order primitive update with small characteristic-based dissipation transports sharp real-gas gradients cleanly while preserving the constant velocity and pressure fields.

\section{Conclusion} \label{sec:conclusion}
We proposed the Central Differential flux with High-Order Dissipation (CDHD) to address spurious oscillations and energy conservation errors in transcritical flow simulations. The method combines a central flux with minimal upwind-biased dissipation, achieving second-order accuracy in smooth regions while substantially improving energy conservation. Embedded in a hybrid framework with a conservative shock-capturing scheme, CDHD demonstrates robust performance across both smooth and shock-containing flows. These results establish CDHD as a reliable and efficient approach for real-gas simulations in engineering and scientific applications.


\printcredits

\bibliographystyle{unsrt}

\bibliography{cas-refs}


\end{document}